\documentclass[prl,twocolumn,showpacs]{revtex4} 
\usepackage{ae}
\usepackage{graphicx} 

\newcommand{\rb}{$^{87}$Rb}
\newcommand{\rbo}{$^{85}$Rb}
\newcommand{\kq}{$^{41}$K}

\newcommand{\kqa}{$^{40}$K}

\renewcommand{\prl}{Phys.~Rev.~Lett.}
\renewcommand{\pra}{Phys.~Rev.~A}

\renewcommand{\rmp}{Rev.~Mod.~Phys.}

\newcommand{\om}{2\pi f_{\rm m}}

\begin{document}

\title{Association of ultracold double-species bosonic molecules}

\author{C.~Weber$^{1,*}$}
\author{G.~Barontini$^{1}$}
\author{J.~Catani$^{1,2}$}
\author{G.~Thalhammer$^{1}$}
\author{M.~Inguscio$^{1,2}$}
\author{F.~Minardi$^{1,2}$}

\affiliation{$^1$LENS - European Laboratory for Non-Linear
  Spectroscopy and Dipartimento di Fisica, Universit\`a di Firenze,
  via N. Carrara 1, I-50019 Sesto Fiorentino - Firenze, Italy\\
  $^2$CNR-INFM, via G. Sansone 1, I-50019 Sesto Fiorentino -
  Firenze, Italy\\
  $^*$Institut f$\ddot{u}$r Angewandte Physik, Universit$\ddot{a}$t
  Bonn, Wegelerstra\ss e 8, D-53115 Bonn, Germany.}

\begin{abstract}
  We report on the creation of heterospecies bosonic molecules,
  associated from an ultracold Bose-Bose mixture of \kq\ and \rb, by
  using a resonantly modulated magnetic field close to two Feshbach
  resonances. We measure the binding energy of the weakly bound
  molecular states versus the Feshbach field and compare our results
  to theoretical predictions. We observe the broadening and asymmetry
  of the association spectrum due to thermal distribution of the
  atoms, and a frequency shift occurring when the binding energy
  depends nonlinearly on the Feshbach field. A simple model is
  developed to quantitatively describe the association process. Our
  work marks an important step forward in the experimental route
  towards Bose-Einstein condensates of dipolar molecules.
\end{abstract}


\date{\today}

\maketitle

Ultracold polar molecules hold the promise of a revolution in the
domain of quantum degenerate gases and precision measurements.
Degenerate molecules are sought primarily to produce a gas with
strong long-range interactions, stemming from the coupling of
electric dipole moments that heterospecies dimers feature. Such
molecules would create strongly correlated systems with a wealth
of quantum phases \cite{goral_dipolar}, provide candidate qubits
\cite{demille-quantumcomp-PRL02, micheli-nature}, allow for a new
generation of dipolar Bose-Einstein condensates (BECs)
\cite{lewe-dipolarBEC-PRL00}, and help in the search of the
electron dipole moment \cite{kozlov-electron}. Starting from
ultracold atoms, molecules have been successfully created
following two different approaches: photoassociation
\cite{heinzen-photoass} and magnetoassociation \cite{magnetoass},
\begin{figure}[b]
\centering
\includegraphics[width=0.95\columnwidth]{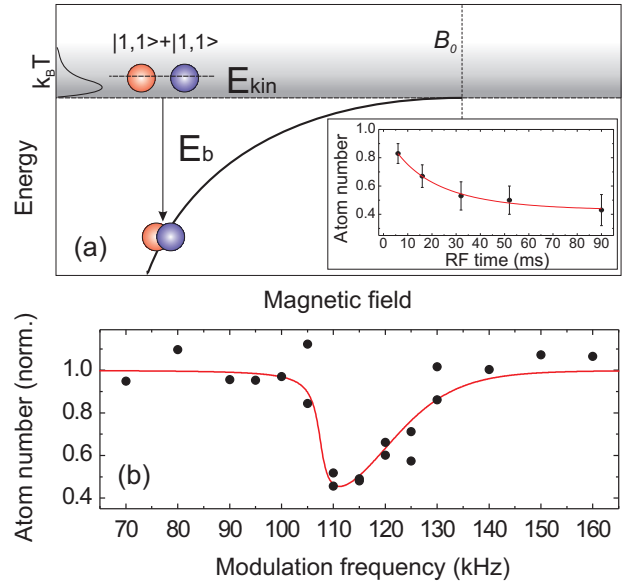}
\caption{(Color online): (a) Molecular radio-frequency association:
  two free atoms with relative kinetic energy $E_{\rm kin}$ are
  resonantly transferred to a molecular level by modulating the
  magnetic field. The inset shows the remaining atom number after a
  variable association time. The solid line is an non-exponential fit
  to data (see text).  (b) Spectrum for an association time
  of 30\,ms, $B$=78.25\,G. The solid line is a fit with the lineshape
  described in text.}
\label{fig:cartoon}
\end{figure}
but few experiments have hitherto reported the production of Feshbach
heteronuclear dimers. Two groups have reported the creation of
fermionic KRb molecules \cite{sengstock-4087-PRL06, jin-RF-PRA08}, while
the only bosonic dimer so far associated is \rbo\rb\
\cite{wieman-8587-PRL05}, which can not be dipolar since the
constituents share the same electronic configuration. Heterospecies
bosonic dimers, i.e., the constituents of the dipolar BEC envisioned in
Ref.~\cite{lewe-dipolarBEC-PRL00}, have instead eluded experimental
realization so far.

A Bose-Bose mixture is particularly suitable to associate such dimers
due to the high phase-space densities achievable, while the atom-dimer
relaxation, which limits the lifetime of the molecules, can be
strongly suppressed in optical lattices with a single atom pair per
lattice site \cite{grimm-mottMol-PRL06}.  Recent progresses in
molecular stabilization schemes \cite{deepbound} make the Bose-Bose
\kq\rb\ mixture truly promising for the experimental observation of
BECs of dipolar molecules. Following a different route, other
experiments have very recently obtained ultracold heterospecies
Fermi-Fermi mixtures \cite{schreck-lik-PRL08, dieckmann-lik-PRL08}
that also can provide a way to compound bosonic dimers.

In this Letter, we report on the production of heterospecies \kq\rb\
bosonic molecules starting from an ultracold mixture. In proximity of
Feshbach resonances (FR's) at moderate magnetic fields, by adding a
modulation to the Feshbach field \cite{wieman-85-PRL05}, we have
converted up to 12\,000 \kq\rb\ pairs into dimers, i.e., 40\% of the
minority \kq\ atoms, at temperatures between 200 and 600\,nK. We
estimate the molecular lifetime to be at least 60\,$\mu$s at atomic
densities of $\sim 5\times 10^{11}$\,cm$^{-3}$ for each species.

\begin{figure}[t]
\centering
\includegraphics[width=\columnwidth]{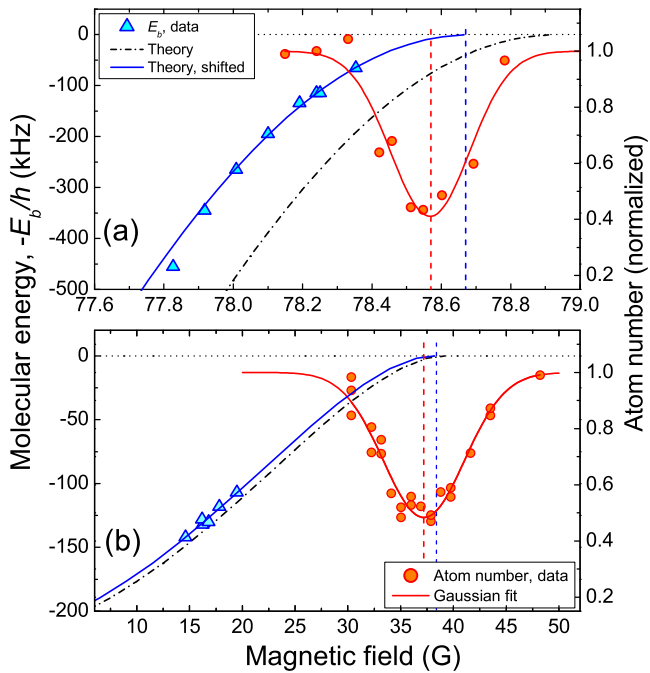}
\caption{(Color online): (a) KRb binding energy and 3-body losses for
  the Feshbach resonance at 79\,G: triangles show the experimental
  binding energies, circles the atom number. The dash-dotted black
  line shows the theoretical prediction for $E_{\rm b}$, the solid
  blue line is the same curve but translated by $-0.24\,$G.  The solid
  red line represents a Gaussian fit to the atom number.  The vertical
  dashed lines show the position of the FR as determined by the 3-body
  losses (red, left) and the binding energy (blue, right). (b) Same as
  above, for the FR at 39\,G. Note the widely different scale of
  $x$-axis. Here the theoretical model for $E_{\rm b}$ is shifted by
  $-1\,$G.}
\label{fig:eb}
\end{figure}

As described earlier \cite{desarlo-k39-PRA07}, we start from a
double-species MOT of \kq\ and \rb\ loaded by two separate 2D MOT's
\cite{catani-2dmot-PRA06}. A quadrupole magnetostatic trap moves the
atoms to the center of a millimetric Ioffe trap, that confines the
sample during microwave evaporation on the \rb\ hyperfine transition
and allows for sympathetic cooling between Rb and K. At a temperature
of 1.5\,$\mu$K, we load an optical trap created by two orthogonal laser
beams ($\lambda$=1064\,nm, waist=110\,$\mu$m) and transfer \rb\ and
\kq\ from the $|F=2,m=2\rangle$ to the $|1,1\rangle$ state by microwave
and radio-frequency adiabatic passage, respectively, in presence of a
7\,G bias magnetic field.  All \rb\ atoms left in $|2,2\rangle$
($\sim$\,10\%) are removed by a pulse of resonant light. No blast light
is needed for \kq\ atoms. At this stage we turn on a uniform magnetic
Feshbach field of approximately 78\,G, close to the previously
observed FR \cite{gregor-dbec-PRL08}, corresponding to
a value of the interspecies scattering length $a \sim200\,a_0$. By
lowering the dipole trap power, we further evaporate the gas down to a
temperature as low as 200\,nK. At this temperature we have
approximately $4(3)\times 10^4$ atoms of \rb\ (\kq).

To associate KRb molecules we bring the Feshbach field to the desired
value below the resonance position $B_0$ and modulate it by adding a
radio-frequency (RF) field of typical amplitude $\sim$130\,mG, as
depicted in Fig.~\ref{fig:cartoon}(a). This modulation pulse has a
simple square envelope, with a typical duration of 20\,ms. At magnetic
fields $B$ lower than the resonance value $B_0$, dimers are created
when the modulation frequency $f_{\rm m}$ is close to the binding
energy $E_{\rm b}/h$ of the molecular level. The production of
molecules is revealed by the reduction of the total number of atoms $N
= N_{\rm K} + N_{\rm Rb}$ at the end of the RF pulse, as we scan the
modulation frequency. A typical lineshape representing the
molecular association peak is reported in
Fig.~\ref{fig:cartoon}(b). Our molecules decay, most likely because of
vibrational quenching induced by collisions with unpaired atoms~\cite{jin-RF-PRA08}, and
leave the trap releasing the acquired binding energy.

For several values of $B$, we measure the corresponding binding energy
by fitting the experimental RF spectra with the lineshape model
described later: the results are shown in Fig.~\ref{fig:eb} together
with the theoretical predictions of the collisional model
\cite{simoni-model}.  For both FR's the agreement is fairly good,
provided the theoretical curve is translated by $-0.24$\,G
[Fig.~\ref{fig:eb}(a)] and $-1$\,G [Fig.~\ref{fig:eb}(b)] along the
$B$ field axis.  A precise determination of the position of the FR's
is given by the points where the molecular energy curves
fitting our data cross the $E_{\rm b}$=0 threshold: 38.2\,G and
78.67\,G. To stress the sensitivity of this method, it is worth to
remark that a deviation of 50\,mG in the $B$ field leads to a shift of
$\sim$50\,kHz in the binding energy of the leftmost experimental point
of Fig.~\ref{fig:eb}(a). In our experiment, the Feshbach field
stability represents the main limitation on the precision of this
method. We calibrate the Feshbach field from the frequency of Rb
hyperfine transitions. Given the fluctuations observed during a single
measurement and day-to-day, we associate an uncertainty of $\pm
30$\,mG to the $B$ field values.

We compare these FR positions with the determination obtained by the
3-body losses. We have repeated the loss measurements of
Ref.~\cite{gregor-dbec-PRL08}, the data are also shown in
Fig.~\ref{fig:eb}. For both FR's, the position obtained by the peak of
the 3-body losses lays slightly below the $B$ value obtained by the
binding energy measurements. Actually, a more comprehensive analysis
of the dynamics associated with 3-body losses is required to extract
the FR position, especially for the the low-field broad FR.  We also
remind that the extrapolation to zero binding energy is performed
under the assumption that the theoretical predictions of
Ref.~\cite{simoni-model} are accurate except for a small shift in the
$B$ field.

Our experimental RF spectra display several interesting and
non-trivial features: {\it (i)} a stark asymmetry and a pronounced
broadening related to the finite temperature of the atoms, {\it (ii)}
a shift of the resonant modulation frequency that increases with the
modulation amplitude, and {\it (iii)} additional association peaks at
fractional frequencies of the binding energy. A precise description of
the above mentioned features is crucial to accurately determine the
binding energies from the measured RF spectra.

The physical reason of the asymmetric lineshape shown in
Fig.~\ref{fig:cartoon}(b) is easily understood: as discussed later,
the resonant modulation frequency of a given atom pair depends on the
pair kinetic energy $E_{\rm kin}$, therefore the spectrum is
inhomogeneously broadened by the asymmetric Boltzmann distribution of
kinetic energies $\sqrt{E_{\rm kin}}\exp(-E_{\rm kin}/k_{\rm
  B}T)$~\cite{hanna-pra07}. Indeed for our typical temperatures, that
is 200--600\,nK, thermal broadening is the dominating contribution to
the observed linewidth. The asymmetry went unobserved in related
experiments of molecular association \cite{wieman-85-PRL05,
  jin-RF-PRA08} that employ degenerate or nearly degenerate gases.

In the following we describe the simple model developed to
analyze our data by combining ideas from the theoretical description
of photoassociation \cite{Mackie2000} with a model of the dynamics of
molecule association in a harmonic oscillator well
\cite{Bertelsen2007}.
Our model is based on several simplifying assumptions. In the
center-of-mass frame, we restrict to three quantum levels: one for
each atomic species in the continuum and another one for the molecular
bound state. Similar to \cite{hanna-pra07} the multitude of continuum
states is taken into account by the thermal average of the results.
We start with a set of nonlinear differential equations for the
 amplitudes $a_j$ ($j=1,2$) and $m$ of the two atomic and
the molecular state, respectively:
\begin{eqnarray}
  i\dot a_j &=& \Omega \cos(\om t) a_{k}^\ast m \quad k\neq j  \label{eq:1}\\
  i\dot m &=& \Omega \cos(\om t) a_1a_2  - \left(E(t)/\hbar + i\gamma/2\right) m
  \nonumber
\end{eqnarray}
Here $E(t) = E_{\rm kin} +E_{\rm b}(t)$, $E_{\rm kin}$ is the relative
kinetic energy of the atoms in their center-of-mass frame, $E_{\rm
  b}(t) = \eta [\Delta B + B_{\rm m} \sin (\om t)]^2$ the
time-dependent binding energy at an average detuning $\Delta B$ from
the FR. We use the quadratic form of $E_{\rm b}$, with curvature
$\eta$, valid in the universal regime, and
include the time variation due to the modulation of the magnetic
field.
For reasons of simplicity, we assume that the coupling strength
$\Omega$, proportional to the modulation amplitude $B_{\rm m}$, is the
same for all states under consideration and that the molecule decay
rate $\gamma$ is independent of atomic and molecular densities.

\begin{figure}[t]
\centering
\includegraphics[width=\columnwidth]{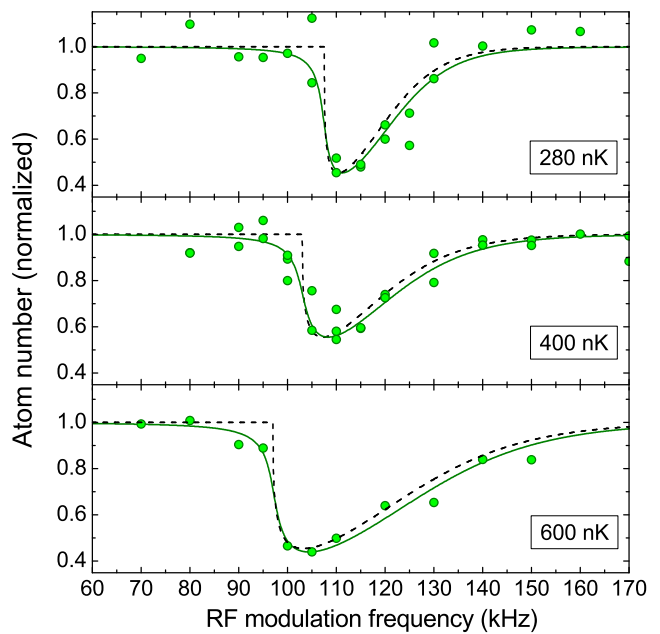}
\caption{(Color online): Lineshapes of the molecular association
  process for different values of temperature. Full circles represent
  data, taken after 30\,ms of RF pulse length around $B$=78.2\,G. Solid
  lines represents the fitted lineshape with the model described in
  the text: we take the binding energy $E_{\rm b}$, the decay rate $\gamma$
  and the coupling strength $\Omega$ as free fit parameters. The
  temperature is measured from the ballistic expansion of the
  clouds. Dashed lines show the fits assuming
  $\gamma=0$.} \label{fig:temperature}
\end{figure}

Numerical solutions of Eq.~\ref{eq:1} describe several key features of
our experimental observations:
For sufficiently small modulation amplitudes this model predicts atom
loss when the modulation frequency equals $f_0 \equiv (E_{\rm kin} +
E_{\rm b})/h$. The total atom number $N(t)=|a_1(t)|^2 + |a_2(t)|^2$
decays non-exponentially: for balanced atomic populations, we find
\begin{equation}
N(f_{\rm},t) = \frac{N(0)}{1 + N(0)\,\kappa(f_{\rm m})\, t}
  \nonumber
  \label{eq:2}
\end{equation}
in the limit $\Omega \ll \gamma$. In our measurements we observe such
an non-exponential decay, as shown in the inset of
Fig.~\ref{fig:cartoon}. This corresponds with the solution of the rate
equation $\dot N = -\kappa(f_{\rm m}) N^2$ which is characteristic for
molecule formation as a two-body process. The calculated atom loss
coefficient $\kappa(f_{\rm m})$ shows a Lorentzian dependence on the
modulation frequency $f_{\rm m}$, with center and width given by $f_0$
and $\gamma$. In order to account for the thermal distribution of the
kinetic energies, we still fit our spectra with Eq.~\ref{eq:2},
after replacing $\kappa(f_{\rm m})$ by its convolution with
the Boltzmann distribution: the results well describe our data, as
shown by the fits in Fig.~\ref{fig:temperature}.

Together with the binding energy, from fits we obtain an estimate for
the molecular lifetime $\tau=1/\gamma=60\,\mu$s with an uncertainty of
a factor 3. For comparison we report in Fig.~\ref{fig:temperature}
also the fit assuming no decay: it is evident that the finite $\gamma$
smoothens the left edge of the lineshape. Actually, from the spectra
we can only infer a lower limit for the lifetime, that is
$\tau>20\,\mu$s, but not an upper limit. Indeed the observed
smoothening could be also explained by technical reasons, for example
magnetic field fluctuations during the RF pulse. Instead, an upper
limit $\tau<5$\,ms is deduced from the lack of an observable atom
decay after the RF pulse is terminated.

\begin{figure}[t]
\centering
\includegraphics[width=\columnwidth]{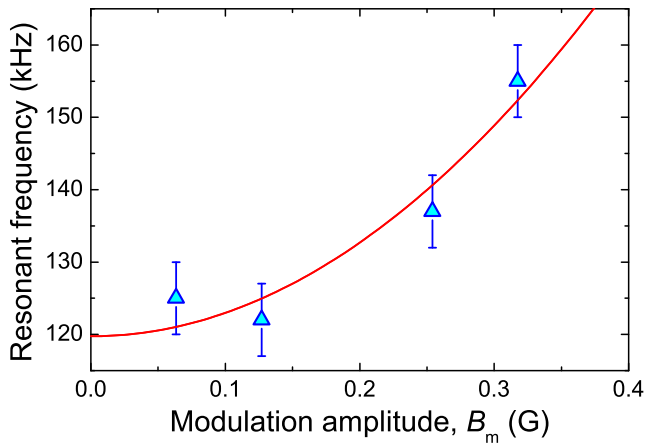}
\caption{(Color online): Shift of the resonant modulation frequency as
  a function of the modulation amplitude, measured at $B$=78.30\,G
  with 15\,ms of RF pulse length and a temperature of 350\,nK. The
  solid line represents the parabolic fit.}
\label{fig:amplitude}
\end{figure}

As mentioned above, other non-trivial features appear at high
modulation amplitudes, that are also described by our model.
First, we observe a shift of the modulation resonant frequency,
increasing with the modulation amplitude (see
Fig.~\ref{fig:amplitude}), that can be understood in a very direct way
considering the quadratic dependence of the binding energy on the
magnetic field: the time-averaged value of the transition energy
$\langle E(t) \rangle$ in Eq.~\ref{eq:1} deviates from the value
$E_{\rm kin} +\eta \Delta B^2$ by an amount $\eta B_{\rm
  m}^2/2$. We directly verified this result by measuring the resonant
modulation frequency as a function of the modulation amplitude $B_{\rm
  m}$ (see Fig.~\ref{fig:amplitude}). As a consequence, for the
high-field FR we extrapolate to zero modulation amplitude and correct
the measured resonant frequencies by $-6$\,kHz, corresponding to our
modulation amplitude of 0.13\,G. For the low-field FR, no shift due to
the amplitude modulation occurs since the molecular level is
approximately linear with the magnetic field.

Moreover, we observe additional loss peaks occurring when the
modulation frequency is at fractional values of the average transition
energy $h f_{\rm m} = \langle E(t)\rangle/n$ with integer $n$. This
allows us to access a range of binding energies which lies outside the
useable bandwidth of our excitation coil. The data points at large
binding energy of Fig.~\ref{fig:eb} are deduced from such measurements
with a modulation frequency corresponding to half the transition
energy.

Finally, we comment on the markedly different association efficiency
observed between the two FR's. On the low-field Feshbach resonance the
dimers association shows a much poorer efficiency and requires an RF
pulse of 1\,s at maximum modulation amplitude 0.5\,G. This can be
understood considering that the coupling strength $\Omega$ of
Eq.~\ref{eq:1} is significant only at field detunings $\Delta B$ where
the bare states of the open and closed channel mix and the energy of
the molecular level deviates from the linear dependence on the $B$
field~\cite{feshbach-rmp07-julienne}. As shown in Fig.~\ref{fig:eb},
this is clearly the case only for the Feshbach resonance around
79\,G. Indeed we calculate that, in the range of our measurements,
$\Omega$ is approximately a factor 10 smaller on the low-field FR,
also taking into account the larger modulation amplitude employed. The
reduction of association rate by two order of magnitudes is reflected
in the need of an elongated RF pulse.

In conclusion, we have reported the creation of heterospecies \kq \rb\
ultracold molecules, representing the first feasible starting point
toward a BEC of dipolar molecules, since the different electronic
structure of the constituents allows for a permanent electric dipole
moment in the vibrational ground state. On the most favorable FR, with
thermal clouds at 200\,nK we converted up to 40\% of the minority
atoms. We carried out a detailed yet simple analysis of the
association, highlighting a remarkable shift of the atoms-molecule
transition that must be taken into account for a proper determination
of the binding energy, when the latter depends nonlinearly on the
Feshbach field.

We plan to confine the atoms in an optical lattice
\cite{catani-3dlat-PRA08} to suppress the molecular decay due to
inelastic collisions. This should grant enough time to drive the
molecules to a lower -- possibly the ground -- vibrational
states. Last but not least, molecular association will also serve as
an important tool in the exploration of the quantum phases of the
Bose-Bose mixture in the optical lattice.

This work was supported by Ente CdR in Firenze, CNR under project
EuroQUAM, EU under STREP NAME-QUAM, and INFN through project
SQUAT-Super.  We thank A. Simoni for sharing his theoretical results,
and G.~Varoquaux who contributed to the experiment, and our collegues
of the Quantum Degenerate Gases group at LENS for fruitful
discussions.

\textit{Note added.} After the completion of this work, we have become
aware of the production of bosonic molecules from the Fermi-Fermi
mixture $^6$Li\kqa\ \cite{dieckmann-icap}.


\begin{thebibliography}{99}

\bibitem{goral_dipolar}
K.~Goral, L.~Santos, and M.~Lewenstein, \prl\
{\bf 88}, 170406 (2002).

\bibitem{demille-quantumcomp-PRL02}
D.~DeMille, \prl\ {\bf 88}, 067901 (2002).

\bibitem{micheli-nature}
A.~Micheli, G.~K.~Brennen, and P.~Zoller, Nature Physics {\bf
2}, 341 (2006).

\bibitem{lewe-dipolarBEC-PRL00} L.~Santos, G.~V.~Shlyapnikov,
  P.~Zoller, and M.~Lewenstein, \prl\ {\bf 85}, 1791 (2000).

\bibitem{kozlov-electron}
M.~G.~Kozlov and L.~N.~Labzowsky, J.\ Phys.\ B: At.\ Mol.\
Opt.\ Phys.\ {\bf 28}, 1933 (1995).

\bibitem{heinzen-photoass} R.~H.~Wynar, R.~S.~Freeland, D.~J.~Han,
  C.~Ryu, and D.~J.~Heinzen, Science \textbf{287}, 1016 (2000).

\bibitem{magnetoass} E. A. Donley et al., Nature \textbf{417}, 529
  (2002); F.~A.~van Abeelen and B.~J.~Verhaar,
  Phys. Rev. Lett. \textbf{83}, 1550 (1999); F.~H.~Mies,
  E.~Tiesinga, P.~S.~Julienne, Phys. Rev. A \textbf{61}, 022721
  (2000).

\bibitem{sengstock-4087-PRL06} C.~Ospelkaus, S.~Ospelkaus, L.~Humbert,
  P.~Ernst, K.~Sengstock, and K.~Bongs, \prl\ {\bf 97}, 120402 (2006).

\bibitem{jin-RF-PRA08} J.~J.~Zirbel {\it et al.},
\prl\ {\bf 100}
  143201 (2008); J.~J.~Zirbel {\it et al.}, 
\pra\ A {\bf 78}, 013416 (2008).

\bibitem{wieman-8587-PRL05} S.~B.~Papp, and C.~E.~Wieman, \prl\ {\bf
    97}, 180404 (2006).

\bibitem{grimm-mottMol-PRL06} G.~Thalhammer, K.~Winkler, F.~Lang,
  S.~Schmid, R.~Grimm, and J.~H.~Denschlag, \prl\ {\bf 96},
  050402 (2006).

\bibitem{deepbound} S.~Ospelkaus {\it et al.},
Nature Physics {\bf 4}, 622 (2008); J.~G.~Danzl {\it et al.}, 
Science {\bf 321}, 1062 (2008).

\bibitem{schreck-lik-PRL08}E.~Wille {\it et al.}, 
\prl\ {\bf 100}, 053201 (2008).

\bibitem{dieckmann-lik-PRL08} M.~Taglieber, A.~-C.~Voigt, T.~Aoki,
  T.~W.~H\"ansch, and K.~Dieckmann, \prl\ {\bf 100}, 010401 (2008).

\bibitem{wieman-85-PRL05} S.~T.~Thompson, E.~Hodby, and C.~E.~Wieman,
\prl~{\bf 95}, 190404 (2005).

\bibitem{desarlo-k39-PRA07} L.~De Sarlo, P.~Maioli, G.~Barontini,
J.~Catani, F.~Minardi, and M.~Inguscio, \pra\ {\bf 75},
022715(2007).

\bibitem{catani-2dmot-PRA06} J.~Catani, P.~Maioli, L.~De Sarlo, F.~Minardi,
and M.~Inguscio, \pra\ {\bf 73}, 033415 (2006).

\bibitem{gregor-dbec-PRL08} G.~Thalhammer, G.~Barontini, L.~De Sarlo,
J.~Catani, F.~Minardi, and M.~Inguscio, Phys.\ Rev.\ Lett.\
{\bf 100}, 210402 (2008).

\bibitem{simoni-model} A.~Simoni (private communication); A.~Simoni {\it et
  al.}, \pra\ {\bf 77}, 052705 (2008).

\bibitem{hanna-pra07} T.~M.~Hanna, T.~K\"ohler, and K.~Burnett, \pra\ {\bf
    75}, 013606 (2007).

\bibitem{Mackie2000} M.~Mackie, R.~Kowalski, and J.~Javanainen,
  Phys.~Rev.~Lett. {\bf 84}, 3803 (2000).

\bibitem{Bertelsen2007} J.~F.~Bertelsen and K.~M\"olmer, Phys.~Rev.~A
  {\bf 76}, 043615 (2007).

\bibitem{feshbach-rmp07-julienne} T.~K\"ohler, K. G\'oral, and
  P.~S.~Julienne, \rmp\ {\bf 78}, 1311 (2006).

\bibitem{catani-3dlat-PRA08} J.~Catani, L.~De Sarlo, G.~Barontini,
  F.~Minardi, and M.~Inguscio, \pra\ {\bf 77}, 011603(R) (2008).

\bibitem{dieckmann-icap} A.-C.~Voigt, M.~Taglieber, L.~Costa, T.~Aoki,
  T.~W.~H\"ansch, and K.~Dieckmann, poster at International Conference on
  Atomic Physics XXI, ICAP 2008.

\end{thebibliography}
\end{document}